\documentclass[twocolumn,prb,amsmath,amssymb,floatfix]{revtex4}

\usepackage{bbold}
\usepackage{bm}
\usepackage{color}
\usepackage{dcolumn}
\usepackage{feynmf} 
\usepackage{graphics}
\usepackage{graphicx}
\usepackage{subfigure}
\usepackage{slashed}
\usepackage{placeins}

\usepackage{epsfig}

\def\al{\alpha}

\def\veps{\varepsilon}

\newcommand{\ep}{\varepsilon}

\def\be{\begin{equation}}
\def\ee{\end{equation}}
\def\bea{\begin{eqnarray}}
\def\eea{\end{eqnarray}}
\def\bse{\begin{subequations}}
\def\ese{\end{subequations}}
\def\bc{\begin{center}}
\def\ec{\end{center}}

\def\ra{\rightarrow}

\def\nonum{\nonumber}

\def\Ord{{\rm O}}

\begin{document}

\begin{fmffile}{fmfdymN}
\fmfcmd{%
vardef cross_bar (expr p, len, ang) =
((-len/2,0)--(len/2,0))
rotated (ang + angle direction length(p)/2 of p)
shifted point length(p)/2 of p
enddef;
style_def crossed expr p =
cdraw p;
ccutdraw cross_bar (p, 5mm,  45);
ccutdraw cross_bar (p, 5mm, -45)
enddef;}

\title{Critical behaviour of ($2+1$)-dimensional QED: \\
       $1/N_f$-corrections in the Landau gauge}

\author{A.~V.~Kotikov$^1$, V.~I.~Shilin$^1$ and S.~Teber$^{2,3}$}
\affiliation{
$^1$Bogoliubov Laboratory of Theoretical Physics, Joint Institute for Nuclear Research, 141980 Dubna, Russia.\\
$^2$Sorbonne Universit\'es, UPMC Univ Paris 06, UMR 7589, LPTHE, F-75005, Paris, France.\\
$^3$CNRS, UMR 7589, LPTHE, F-75005, Paris, France.}

\date{\today}

\begin{abstract}
The dynamical generation of a fermion mass is studied within ($2+1$)-dimensional QED
with $N$ four-component fermions in the leading and next-to-leading
orders of the $1/N$ expansion. The analysis is carried out in the Landau gauge which is supposed to insure the gauge independence
of the critical fermion flavour number, $N_c$. It is found that the dynamical fermion mass
appears for $N<N_c$ where $N_c=3.29$, that is only about $1\%$ larger than its value at leading order.
\end{abstract}

\maketitle


\section{Introduction}

Quantum Electrodynamics in $2+1$ dimensions (QED$_3$) has been extensively studied during more than three decades now.
Originally, the interest in QED$_3$ came from its similarities to ($3+1$)-dimensional QCD and the fact that phenomena such as dynamical 
chiral symmetry breaking (D$\chi$SB) and mass generation may be studied systematically in such a toy model, see, {\it e.g.}, 
Refs.~[\onlinecite{Pisarski84,AppelquistNW88,Pennington91+92,Pisarski91,AtkinsonJM90,DagottoKK89+90,Azcoiti93+96,KarthikN16,AppelquistW04,AppelquistCS99,GiombiKT16,DiPietroKSS16,Nash89,Kotikov93+12,BashirR07,BashirRSR09}].
Later, a strong interest in QED$_3$ arose in connexion with planar condensed matter physics systems
having relativistic-like low-energy excitations such as some two-dimensional antiferromagnets~\cite{MarstonA89+IoffeL89}
and graphene;~\cite{Semenoff84+Wallace47} the study of a dynamically generated gap in the fermion spectrum of graphene
has now become an active area of research, see, {\it e.g.}, the reviews Refs.~[\onlinecite{KotovUPGC12,MiranskyS15}].
In all cases, the understanding of the phase structure of QED$_3$ is a crucial pre-requisite to understand 
non-perturbative dynamic phenomena in more realistic particle and condensed matter physics models.

Despite the fact that a large number of investigations have been carried out to study D$\chi$SB 
in QED$_3$, very different results have been obtained. 
Without being extensive, let us indeed recall that, in his seminal paper~\cite{Pisarski84}, Pisarski solved the Schwinger-Dyson (SD) gap
equation using a leading order (LO) $1/N$-expansion and found that a fermion mass is generated for all values of $N$, decreasing exponentially with $N$ and vanishing only in the limit $N \ra \infty$.
Later, he confirmed his finding by a renormalization group analysis.~\cite{Pisarski91}
Support of Pisarski's result was given by Pennington and collaborators~\cite{Pennington91+92} who adopted
a more general non-perturbative approach to solving the SD equations.
On the other hand, in a more refined analysis of the gap equation at LO of the $1/N$-expansion, 
Appelquist et al.~\cite{AppelquistNW88} have shown that the theory exhibits a critical
behaviour as the number $N$ of fermion flavours approaches $N_c = 32/ \pi^2$;
that is, a fermion mass is dynamically generated only for $N<N_c$. 
Contrary to all previous results, an alternative non-perturbative study by
Atkinson et al.~\cite{AtkinsonJM90} suggested that chiral symmetry is unbroken at
sufficiently large $N$. The theory has also been simulated on the lattice.~\cite{DagottoKK89+90,Azcoiti93+96,KarthikN16}
Remarkably, the conclusions of Ref.~[\onlinecite{DagottoKK89+90}] are in the agreement
with the existence of a critical $N$ as predicted in the analysis of Ref.~[\onlinecite{AppelquistNW88}]
while the second paper~\cite{Azcoiti93+96} finds D$\chi$SB for all $N$ and the recent third one~\cite{KarthikN16} no sign of D$\chi$SB at all.  
Even in the case where a finite $N_c$ is found, its value is subject to uncertainty with estimates ranging from $N_c=1$ to $N_c=4$, see
Ref.~[\onlinecite{AppelquistW04}] for a review. Moreover, Ref.~[\onlinecite{AppelquistCS99}] found an upper bound, $N_c <3/2$, 
while, more recently, Ref.~[\onlinecite{GiombiKT16}] found that $N_c < 4.4$ and Ref.~[\onlinecite{DiPietroKSS16}] that $N_c<9/4$.
Clearly, all these disagreements reflect our poor understanding of this problem.

The purpose of the present work is to include $1/N$ corrections to the LO 
result of Ref.~[\onlinecite{AppelquistNW88}].
Because the critical value $N_c$ is not large, the contribution of such higher
orders in the $1/N$ expansion can be essential and their proper study may lead to a better
understanding of the problem. 
This important issue has been rarely addressed in the past. To the best of our knowledge, the main references
are [\onlinecite{Nash89}] and [\onlinecite{Kotikov93+12}] where rather different results
were obtained. 
The well-known results of Ref.~[\onlinecite{Nash89}] demonstrated a quite strong stability
of the $1/N$ expansion while the ones of Ref.~[\onlinecite{Kotikov93+12}] showed that a similar property holds only in the Landau gauge.
The strong gauge dependence found in Ref.~[\onlinecite{Kotikov93+12}]
is in agreement with the studies of Ref.~[\onlinecite{BashirR07}] in the so-called rainbow approximation.
In the following, we shall refine the analysis of Ref.~[\onlinecite{Kotikov93+12}] and perform an accurate computation of all $1/N$ corrections in the Landau gauge,
with a special focus on the most complicated ones, in order to extract the value of $N_c$.

The last years witnessed a strong progress in the study of the gauge dependence of D$\chi$SB 
in various models, see Ref.~[\onlinecite{AhmadCCR16}] as well as references and
discussions therein. The progress is related to the use of the Landau-Khalatnikov-Fradkin 
transformation.~\cite{LandauK55+Fradkin56} In the case of QED$_3$ in the $1/N$-expansion,
the application of this transformation~\cite{BashirRSR09} has revealed the almost complete lack of gauge dependence
for $N_c$. This confirms that we can limit our analysis to the case of the Landau gauge.

\section{Model and Schwinger-Dyson equations} 

The Lagrangian of massless QED$_3$ with $N$ flavours of fermions reads
\be
L = \overline \Psi ( i \hat \partial - e \hat A ) \Psi - \frac{1}{4} F_{ \mu \nu}^2\, ,
\label{L-QED3}
\ee
where $ \Psi$ is taken to be a four component complex spinor.
In the massless case, which we are considering, the model contains infrared divergences.
The latter soften when the model is analysed in a $1/N$ expansion.~\cite{AppelquistP81,JackiwT81+AppelquistH81}
Since the theory is super-renormalizable, the mass scale is given by the dimensionful coupling constant: $a = Ne^2/8$,
%
%
 which is kept fixed as $N \rightarrow \infty$.
In the four component case, we can introduce the matrices $\gamma_3$ and $\gamma_5$
which anticommute with $\gamma_0$, $\gamma_1$ and $\gamma_2$.
Then, the massless case is invariant under the transformations:
$\Psi \rightarrow \exp(i \alpha_1 \gamma_3) \Psi$ and
$ \Psi \rightarrow \exp(i \alpha_2 \gamma_5) \Psi$. 
Together with the identity matrix and $[\gamma_3, \gamma_5 ]$, we have a $U(2)$ symmetry for each spinor and the
full global ``chiral'' (or rather flavour) symmetry is $U(2N)$. A mass term will break this symmetry
to $U(N) \times U(N)$. It is the dynamical generation of such a mass that we shall consider in the following. 
It is also possible to include a parity non-conserving mass, see for
example Ref.~[\onlinecite{AppelquistBKW86}], but we will not consider this possibility here.

Following Ref.~[\onlinecite{AppelquistNW88}], we now study the solution of the SD equation.
The inverse fermion propagator has the form
\be
S^{-1}(p) = [1 + A(p)]\,\left( i\hat p + \Sigma (p) \right)\, ,
\label{Sigma} 
\ee
where $A(p)$ is the wave-function renormalization and $\Sigma (p)$ is the dynamically generated parity-conserving mass which 
is taken to be the same for all the fermions. Notice that in our definition of $\Sigma (p)$, Eq.~(\ref{Sigma}), the choice of the free vertex
corresponds to the so-called central Ball-Chiu vertex~\cite{BallC80} for the ``more standard''
definition $\tilde{\Sigma} (p) =
\Sigma (p)[1 + A(p)]$.
With these conventions, the SD equation for the fermion propagator may be decomposed into scalar and vector components
as follows:
\begin{subequations}
\label{SD-sigma+A}
\begin{eqnarray}
\tilde{\Sigma}(p)
= \frac{2a}{N} \, \mbox{Tr} \int \frac{d^3 k}{(2 \pi )^3}
\frac{\gamma^{\mu} D_{\mu \nu}(p-k) \Sigma (k) \Gamma^{\nu}(p,k)}
{\left[1 + A(k) \right] \left( k^2 + \Sigma^2(k) \right)}\, ,  
\label{SD-sigma} \\
A(p) p^2 =  -\frac{2a}{N} \, \mbox{Tr} \int \frac{d^3 k}{(2 \pi )^3}
\frac{ D_{\mu \nu}(p-k) \hat p \gamma^{\mu} \hat k \Gamma^{\nu}(p,k)}
{\left[1 + A(k) \right] \left( k^2 + \Sigma^2(k) \right)} \, ,
\label{SD-A}
\end{eqnarray}
\end{subequations}
where $D_{\mu \nu}(p)$ is the photon propagator in the Landau gauge:
\be
D_{\mu \nu}(p) = \frac{g_{\mu \nu} - p_{ \mu} p_{ \nu} / p^2}{p^2 \left[1 + \Pi (p) \right]}\, ,
\label{photon}
\ee
$\Pi(p)$ is the polarization operator and $ \Gamma ^{ \nu}(p,k)$ is the vertex function.
In the following, we shall first consider Eqs.~(\ref{SD-sigma+A}) at the LO approximation
and then study Eq.~(\ref{SD-sigma}) at the NLO level.

\begin{figure}[tl]
    \includegraphics[width=0.2\textwidth]{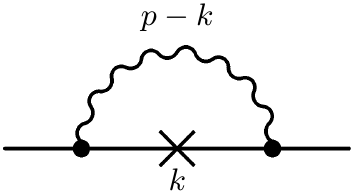}
    \caption{\label{fig:diags-LO}
        LO diagram to the dynamically generated mass $\Sigma(p)$. The crossed line denotes mass insertion.}
\end{figure}

\section{Leading order} 

The LO approximations in the $1/N$ expansion are given by:
\be
A(p) = 0, \quad \Pi (p) = a/ |p|, \quad \Gamma^{\nu}(p,k) = \gamma^{\nu}\, ,
\ee
where the fermion mass has been neglected~\footnote{A study of the fermion mass contribution to $\Pi(p)$ can be
found, for example, in Ref.~[\onlinecite{GusyninHR96}].} in the calculation of $\Pi(p)$. 
A single diagram contributes to the gap equation (\ref{SD-sigma}) at LO, see Fig.~\ref{fig:diags-LO}, and the latter reads:
\be
  \Sigma (p) =\frac{16a}{N}
   \int \frac{d^3 k}{(2 \pi )^3}
\frac{ \Sigma (k) }{ \left( k^2 + \Sigma^2(k)
\right)
\bigl[ (p-k)^2 + a \,|p-k| \bigr]} \, . 
\label{SD-sigma-LO1} 
\ee
Performing the angular integration in Eq.~(\ref{SD-sigma-LO1}) yields:
\be
  \Sigma (p) =\frac{4a}{\pi^2 N |p|}
   \int_0^{\infty} \! \! \! d |k|\, 
\frac{ |k| \Sigma (|k|) }{k^2 + \Sigma^2(|k|)}
\ln \left(\frac{|k|+|p|+a}{|k-p|+a}\right)
\, . 
\label{SD-sigma-LO2} 
\ee
The study of Eq.~(\ref{SD-sigma-LO2}) in Ref.~[\onlinecite{AppelquistNW88}] has revealed the existence
of a critical number of fermion flavours $N_c$, such that for $N>N_c$, $\Sigma (p) =0$. 
As it was argued in this reference, QED$_3$ is strongly damped for $|p|>a$, {\it i.e.}, 
all relevant physics occur at $|p|/a <1$. Hence, only the lowest order terms in $|p|/a$ have to be kept on the r.h.s.\ of Eq.~(\ref{SD-sigma-LO2}) 
with a hard cut-off at $|p|=a$. Moreover, considering $N$ 
close to $N_c$, the value of $\Sigma (|k|)$ can be made arbitrarily small. Thus, 
$k^2 + \Sigma^2(|k|)$ can be replaced by $k^2$ on the r.h.s.\ of Eq.~(\ref{SD-sigma-LO2}) 
which then further simplifies as:
\be
\Sigma (p) =\frac{8}{\pi^2 N}
   \int_0^{a} \! \! \! d |k| \, 
\frac{\Sigma (|k|) }{ \mbox{Max}({|k|,|p|})}\, . 
\label{SD-sigma-LO3} 
\ee
Following Ref.~[\onlinecite{AppelquistNW88}], the mass function may then be parametrized as:
\be
\Sigma (k) = B \, (k^2)^{ -\alpha} \, ,
\label{sigma-parametrization} 
\ee
(with an arbitrary $B$ value) where the index $\al$ has to be self-consistently determined. Substituting (\ref{sigma-parametrization}) in Eq.~(\ref{SD-sigma-LO3}), the gap equation reads:
\bea
1 = \frac{2\beta}{L} \quad \text{where} \quad \beta = \frac{1}{\alpha \left( 1/2 - \alpha \right)} \quad \text{and} \quad  L \equiv \pi^2 N\, .
\label{gap-eqn-LO} 
\eea
Solving the gap equation, the following values of $\al$ are obtained:
\begin{eqnarray}
\alpha_{\pm} = \frac{1}{4}\,\left( 1 \pm \sqrt{1 - \frac{32}{L}} \right) \, ,
\label{al-LO} 
\end{eqnarray}
which reproduces the solution given by Appelquist et al.\ in Ref.~[\onlinecite{AppelquistNW88}]. 
Their analysis yields a critical number of fermions:
$N_c = 32/ \pi^2 \approx 3.24$ ({\it i.e.}, $L_c = 32$),
such that $\Sigma(p) = 0$ for $N>N_c$ and
\be
\Sigma(0) \simeq \exp
\bigl[ -2 \pi / (N_c/N - 1)^{1/2} \bigr]\, ,
\label{Sigma(0)}
\ee
for $N<N_c$. Thus, D$\chi$SB occurs when $\alpha$ becomes
complex, that is for $N<N_c$.

As it was shown in Ref.~[\onlinecite{Kotikov93+12}], the same result for $\Sigma(p)$ can be obtained
in another way. Taking the limit of large $a$, the linearized version of Eq.~(\ref{SD-sigma-LO1})
has the following form:
\be
  \Sigma (p) =\frac{16}{N}
   \int \frac{d^3 k}{(2 \pi )^3} \frac{ \Sigma (k) }{k^2 \, |p-k| } \, . 
\label{SD-sigma-LO4} 
\ee
Interestingly, the large-$N$ limit of the photon propagator in QED$_3$ 
has precisely the same momentum dependence as the one in the so-called reduced QED, see Ref.~[\onlinecite{GorbarGM01}] 
and also Refs.~[\onlinecite{Marino93+DoreyM92+KovnerR90}].
The multi-loop structure of the latter has been recently explored in Refs.~[\onlinecite{Teber12+KotikovT13,KotikovT14}]. 
With the help of the ansatz (\ref{sigma-parametrization}), one can then see that 
the r.h.s.\ of Eq.~(\ref{SD-sigma-LO4}) may be 
calculated with the help of the standard rules of perturbation theory for massless Feynman diagrams 
as in Ref.~[\onlinecite{Kazakov83}], see also the recent short review Ref.~[\onlinecite{TeberK16}].
Indeed, given these rules, the computation of Eq.~(\ref{SD-sigma-LO4}) is straightforward and reads:
\be
\Sigma^{(\text{LO})}(p) = \frac{8B}{N}\,\frac{(p^2)^{-\al}}{(4\pi)^{3/2}}\, \frac{2\beta}{\pi^{1/2}}\, .
\label{sigma-LO-res}
\ee
This  immediately yields the gap equation (\ref{gap-eqn-LO}) and, hence, the results
of Eq.~(\ref{al-LO}) together with the critical value $N_c = 32/ \pi^2 $ at which the index $\alpha$
becomes complex.

Similarly, such rules allow for a straightforward evaluation of the wave function renormalization. At LO, Eq.~(\ref{SD-A}) simplifies as:
\be
  A(p) p^2 = -\frac{2a}{N}
 \text{Tr} \int \! \! \frac{d^D k}{(2 \pi )^D}
\frac{
( g_{\mu \nu} - \frac{(p-k)_\mu (p-k)_\nu}{(p-k)^2})
\hat p
\gamma^{\mu}
\hat k
\gamma^{\nu}}{
k^2 |p-k|}\, , 
\label{SD-A1}
\ee
where the integral has been dimensionally regularized with $D=3 -2 \veps$. 
Taking the trace and computing the integral on the r.h.s.\ yields:
\be
  A(p) =
\frac{\Gamma(1+\ep)(4\pi)^{\ep} \mu^{2\ep}}{p^{2\ep}} \, C_1 = 
  \frac{\overline{\mu}^{2\ep}}{p^{2\ep}} \, C_1 \, + \Ord(\ep) \, ,
\label{SD-A2}
\ee
where the $\overline{MS}$ parameter $\overline{\mu}$ has the standard form
$\overline{\mu}^2 = 4\pi e^{-\gamma_E} \mu^2$ with the Euler constant $\gamma_E$ and
\be
C_1 =
+\frac{4}{3\pi^2N} \left(\frac{1}{\ep} + \frac{7}{3} - 2\ln 2\right)\, .
\ee
The corresponding anomalous scaling dimension of the fermion field then reads: 
$\eta = \mu^2 (d/d\mu^2) A(p) = 4 / (3\pi^{2}N)$,
%
%
and coincides with the one in Ref.~[\onlinecite{Gracey94}].

\section{Next-to-leading order}  

The ease at which the standard rules for computing massless Feynman diagrams allowed us to derive LO results suggests the possibility to extend these computations beyond LO.
We therefore consider the NLO contributions to the dynamically generated mass and parametrize them as:
\be
\Sigma^{(\text{NLO})}(p) = \left(\frac{8}{N}\right)^2 B\,\frac{(p^2)^{-\al}}{(4\pi)^{3}}\,
\left( \Sigma_A + \Sigma_1 + 2\,\Sigma_2 + \Sigma_3 \right) \, ,
\ee
where each NLO contribution is represented graphically in Fig.~\ref{fig:diags-NLO}.
Because we are dealing with the linearized gap equation, each contribution contains a single mass insertion.
Adding these contributions to the LO result, Eq.~(\ref{sigma-LO-res}), 
the gap equation has the following general form:
\begin{eqnarray}
1 = \frac{2\beta}{L} +
\frac{\pi}{L^2}\,\Bigl[\Sigma_A + \Sigma_1 + 2\,\Sigma_2 + \Sigma_3 \Bigr]\, .
\label{gap-eqn-NLO}
\end{eqnarray}
After very tedious and lengthy calculations, all NLO contributions could be evaluated exactly using the rules for computing massless Feynman diagrams.
For the most complicated scalar diagrams, see $I_1(\al)$ and $I_2(\al)$ below, the Gegenbauer-polynomial technique has been used following the paper [\onlinecite{Kotikov95}].
We now summarize our results (details of the calculations will be published elsewhere).

The contribution  $\Sigma_A$, see Fig.~\ref{fig:diags-NLO} A), originates from the LO value of $A(p)$ and is singular.
Using dimensional regularization, it reads:
\be
\overline{\Sigma}_A =
+\frac{16}{3}\, \frac{\overline{\mu}^{2\ep}}{p^{2\ep}}\, \beta \left(\frac{1}{\ep} + \Psi_1 + \frac{4}{3} - \frac{\beta}{4} \right) \, + \Ord(\ep)\, ,
\label{sigma-NLO-A}
\ee
where $\overline{\Sigma}_i = \pi \Sigma_i$, $(i=1,2,3.A)$ and 
\be
\Psi_1 = \Psi(\alpha)+  \Psi(1/2-\alpha)-2\Psi(1) + \frac{3}{1/2-\alpha} 
-2 \ln 2\, ,
\label{psi1}
\ee
and $\Psi$ is the digamma function.
The contribution of diagram 1) in Fig.~\ref{fig:diags-NLO} is finite and reads:
\be
\overline{\Sigma}_1 = -4 \hat{\Pi} \beta, ~~~
\hat{\Pi} = \frac{92}{9}-\pi^2\, ,
\label{sigma-NLO-1}
\ee
where~\cite{GusyninHR01,Teber12+KotikovT13} the contribution of $\hat{\Pi}$ arises from the two-loop polarization operator in dimension $D=3$ which may be graphically represented as:
\be
\parbox{8mm}{
    \begin{fmfgraph*}(8,7)
      \fmfleft{i}
      \fmfright{o}
      \fmfleft{ve}
      \fmfright{vo}
      \fmffreeze
      \fmfforce{(-0.3w,0.5h)}{i}
      \fmfforce{(1.3w,0.5h)}{o}
      \fmfforce{(0w,0.5h)}{ve}
      \fmfforce{(1.0w,0.5h)}{vo}
      \fmffreeze
      \fmf{photon}{i,ve}
      \fmf{photon}{vo,o}
      \fmffreeze
      \fmfdot{ve,vo}
      \fmf{phantom,tag=1}{ve,vo}
      \fmfposition
      \fmfipath{p[]}
      \fmfiset{p1}{vpath1(__ve,__vo)}
      \def\blob#1{%
        \fmfiv{decor.shape=circle,decor.filled=shaded,decor.size=1w}{#1}}
      \blob{point length(p1)/2 of p1}
    \end{fmfgraph*}
} \qquad = \quad 2 \times ~
\parbox{8mm}{
    \begin{fmfgraph*}(14,8)
      \fmfleft{i}
      \fmfright{o}
      \fmf{photon}{i,v1}
      \fmf{photon}{v2,o}
      \fmf{phantom,right,tension=0.1,tag=1}{v1,v2}
      \fmf{phantom,right,tension=0.1,tag=2}{v2,v1}
      \fmf{phantom,tension=0.1,tag=3}{v1,v2}
      \fmfdot{v1,v2}
      \fmfposition
      \fmfipath{p[]}
      \fmfiset{p1}{vpath1(__v1,__v2)}
      \fmfiset{p2}{vpath2(__v2,__v1)}
      \fmfiset{p3}{vpath3(__v1,__v2)}
      \fmfi{plain}{subpath (0,length(p1)) of p1}
      \fmfi{plain}{subpath (0,length(p2)/4) of p2}
      \fmfi{plain}{subpath (length(p2)/4,3length(p2)/4) of p2}
      \fmfi{plain}{subpath (3length(p2)/4,length(p2)) of p2}
      \fmfi{photon}{point length(p2)/4 of p2 .. point length(p3)/2 of p3 .. point 3length(p2)/4 of p2}
      \def\vert#1{%
        \fmfiv{decor.shape=circle,decor.filled=full,decor.size=2thick}{#1}}
      \vert{point length(p2)/4 of p2}
      \vert{point 3length(p2)/4 of p2}
    \end{fmfgraph*}
} \qquad + \quad
\parbox{8mm}{
    \begin{fmfgraph*}(14,8)
      \fmfleft{i}
      \fmfright{o}
      \fmf{photon}{i,v1}
      \fmf{photon}{v2,o}
      \fmf{phantom,right,tension=0.1,tag=1}{v1,v2}
      \fmf{phantom,right,tension=0.1,tag=2}{v2,v1}
      \fmf{phantom,tension=0.1,tag=3}{v1,v2}
      \fmfdot{v1,v2}
      \fmfposition
      \fmfipath{p[]}
      \fmfiset{p1}{vpath1(__v1,__v2)}
      \fmfiset{p2}{vpath2(__v2,__v1)}
      \fmfi{plain}{subpath (0,length(p1)/2) of p1}
      \fmfi{plain}{subpath (length(p1)/2,length(p1)) of p1}
      \fmfi{plain}{subpath (0,length(p2)/2) of p2}
      \fmfi{plain}{subpath (length(p2)/2,length(p2)) of p2}
      \fmfi{photon}{point length(p1)/2 of p1 -- point length(p2)/2 of p2}
      \def\vert#1{%
        \fmfiv{decor.shape=circle,decor.filled=full,decor.size=2thick}{#1}}
      \vert{point length(p1)/2 of p1}
      \vert{point length(p2)/2 of p2}
    \end{fmfgraph*}
} \qquad .
\label{polar-2loops}
\ee
The contribution of diagram 2) in Fig.~\ref{fig:diags-NLO} is again singular. Dimensionally regularizing it yields:
\be
2\,\overline{\Sigma}_2 =
-\frac{16}{3} \,\frac{\overline{\mu}^{2\ep}}{p^{2\ep}} \,\beta \left(\frac{1}{\ep} + \Psi_1 + \frac{7}{3}  + \frac{5\beta}{8} \right)
- 2 \hat{\Sigma}_2  + O(\ep) \, ,
\label{sigma-NLO-2}
\ee
where
\bea
&&\hat{\Sigma}_2 = (1-4\alpha) \beta \Bigl[\Psi'(\alpha) - \Psi'(1/2-\alpha)\Bigl] 
\nonum \\
&&-\frac{\pi}{2\alpha}\,\tilde{I}_1(\alpha) - \frac{\pi}{2(1/2-\alpha)} \tilde{I}_1(\alpha+1) \, ,
\label{sigma-NLO-22}
\eea
and $\Psi'$ is the trigamma function. 
Notice that the singularities in $\overline{\Sigma}_A$ and $\overline{\Sigma}_2$ cancel each other and their sum is therefore finite:
\be
\overline{\Sigma}_A + 2\,\overline{\Sigma}_2 = -\frac{2}{3} \beta \Bigl(7\beta+8 \Bigr) -2 \hat{\Sigma}_2\, .
\ee
This cancellation corresponds to the one of the logarithms, 
$\ln(p/\al)$, in Ref.~[\onlinecite{Nash89}]; the importance of such cancellations
was discussed before, in Ref.~[\onlinecite{AppelquistNW88}]. The dimensionless integral $\tilde{I}_1(\alpha)$ appearing in Eq.~(\ref{sigma-NLO-22}) is defined as:
\bea
&&I_1(\alpha) \equiv \frac{(p^2)^{-\al}}{(4\pi)^3}\, \tilde{I}_1(\alpha)
\label{I1-def} \\
&&= \int \frac{d^3k_1}{(2\pi)^3} \frac{d^3k_2}{(2\pi)^3}
   \frac{1}{|p-k_1|k_1^{2\alpha} (k_1-k_2)^2 (p-k_2)^2|k_2|}\, ,
\nonum
\eea
and obeys the following relation (it can be obtained by analogy with the ones in Ref.~[\onlinecite{Kazakov83}]):
\begin{widetext}

\begin{figure}
    \includegraphics[width=0.99\textwidth]{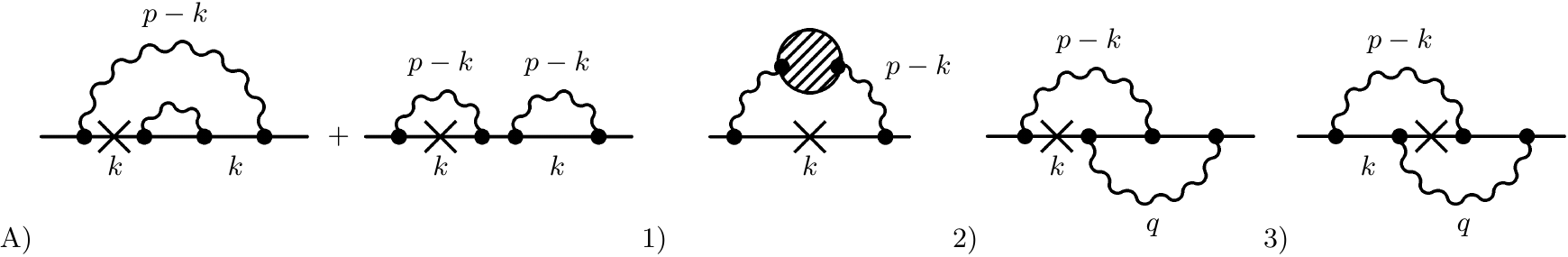}
    \caption{\label{fig:diags-NLO}
        NLO diagrams to the dynamically generated mass $\Sigma(p)$. The shaded blob is defined in Eq.~(\ref{polar-2loops}).}
\end{figure}

\bea
\tilde{I}_1(\alpha+1) =  \frac{(\alpha-1/2)^2}{\alpha^2}
\tilde{I}_1(\alpha) - \frac{1}{\pi \alpha^2}
\Bigl[ \Psi'(\alpha) - \Psi'(1/2-\alpha) \Bigr]\, .
\label{I1-relation}
\eea
Using the results of Ref.~[\onlinecite{Kotikov95}], the integral $\tilde{I}_1(\alpha)$ can be represented
in the form of a two-fold series
\be
\tilde{I}_1(\alpha) = \sum_{n=0}^{\infty} \sum_{l=0}^{\infty}\,\frac{B(l,n,1,1/2)}{(n+1/2)\,\Gamma(1/2)}
\times \Biggl[  \frac{2}{n+1/2} \left(  \frac{1}{l+n+\alpha} +  \frac{1}{l+n+3/2-\alpha} \right) 
 +\frac{1}{(l+n+\alpha)^2} + \frac{1}{(l+n+3/2-\alpha)^2} \Biggr], \,
\label{I1-series}
\ee
where
\bea
B(m,n,\alpha,1/2) = \frac{\Gamma(m+n+\alpha) \Gamma(m+\alpha-1/2)}{m!\Gamma(m+n+3/2)
\Gamma(\alpha)\Gamma(\alpha-1/2)} \, .
\label{B}
\eea
Finally, the contribution of diagram 3) in Fig.~\ref{fig:diags-NLO} is finite and reads:
\bea
\overline{\Sigma}_3 = \hat{\Sigma}_3 + 3 \beta^2, \qquad 
\hat{\Sigma}_3 = 
(1/2-\alpha) \pi \tilde{I}_2(1+\alpha) + \frac{\pi}{2} \tilde{I}_2(\alpha) +
(\alpha-2) \pi \tilde{I}_3(\alpha)\, .
\label{sigma-NLO-3} 
\eea
The dimensionless integrals  in Eq.~(\ref{sigma-NLO-3}) are defined as: $\tilde{I}_2(\alpha)= \tilde{I}(\gamma=1/2,\alpha)$ and 
$\tilde{I}_3(\alpha)= \tilde{I}(\gamma=-1/2,1+\alpha)$, where:
\be
  I(\gamma,\alpha)  \equiv \frac{(p^2)^{-\al -\gamma+1/2}}{(4\pi)^3}\, \tilde{I}(\gamma,\alpha) 
  =\int \frac{d^3k_1}{(2\pi)^3} \frac{d^3k_2}{(2\pi)^3}
   \frac{1}{(p-k_1)^{2\gamma}k_1^{2} (k_1-k_2)^{2\alpha} (p-k_2)^2|k_2|} \, .
   \label{I-def}
\ee
They satisfy the following relations:
\be
 \tilde{I}_2(\alpha) =  \tilde{I}_2(3/2-\alpha),~~~
 \tilde{I}_3(\alpha) = \frac{2}{4\alpha -1} \Bigl(\alpha  \tilde{I}_2(1+\alpha) - (1/2-\alpha)
  \tilde{I}_2(\alpha) \Bigr)
- \frac{\beta^2}{\pi}\, ,
\label{I2-I3-relations} 
\ee
and, thus, only one of them is independent.
Using the results of Ref.~[\onlinecite{Kotikov95}], the integral $\tilde{I}_2(\alpha)$ can be represented
in the form of a
three-fold series:
\begin{subequations}
\label{I2}
\bea
&&\tilde{I}_2(\alpha)
= \sum_{n=0}^{\infty} \sum_{m=0}^{\infty} 
B(m,n,\beta,1/2) \sum_{l=0}^{\infty} 
B(l,n,1,1/2) 
\times  C(n,m,l,\alpha) \, ,
\label{I2-series}\\
&&C(n,m,l,\alpha) =
\frac{1}{(m+n+\alpha)(l+n+\alpha)}
+
 \frac{1}{(m+n+\alpha)(l+m+n+1)}
+  \frac{1}{(m+n+1/2)(l+m+n+\alpha)}
\nonumber \\
&&+ \frac{1}{(m+n+1/2)(l+n+3/2-\alpha)}
 +  \frac{1}{(n+l+\alpha)(l+m+n+\alpha)}
+ \frac{1}{(l+n+3/2-\alpha)(l+n+m+\alpha)}\, .
\label{I2-C}
\eea
\end{subequations}
%

\end{widetext}

Combining all of the above results, the gap equation (\ref{gap-eqn-NLO}) may be written in an explicit form as:
\be
1 = \frac{2\beta}{L} + \frac{1}{L^2}\,
\Bigl[8 S(\al) -\frac{5}{3}\beta^2-\frac{16}{3}\beta -4 \hat{\Pi} \beta  \Bigr]\, ,
\label{gap-eqn-NLO-explicit} 
\ee
where
\be
S(\al) = (\hat{\Sigma}_3(\al)-2 \hat{\Sigma}_2(\al))/8 \, .
\label{delta} 
\ee
At this point, we consider Eq.~(\ref{gap-eqn-NLO-explicit}) directly at the critical point $\alpha=1/4$, {\it i.e.}, at $\beta=16$. 
This yields:
\bea
L_c^2 -32 L_c - 8\bigl(S-64- 8 \hat{\Pi}) = 0 \, ,
 \label{Lc-eqn}
\eea
where $ S= S(\al=1/4)$.
Solving Eq.~(\ref{Lc-eqn}), we have two standard solutions:
\bea
L_{c,\pm} = 16 \pm \sqrt{D},~~~ D= 8(S-32- 8 \hat{\Pi}) \, .
 \label{Lc-solutions}
\eea
It turns out that the ``$-$'' solution is unphysical and has to be rejected because $L_{c,-}<0$. So, the physical solution is unique and corresponds to
$L_{c} = L_{c,+}$.
In order to provide a numerical estimate for $N_c$, we have used the series representations in order to evaluate the integrals: 
$\pi \tilde{I}_1(\alpha=1/4) \equiv R_1$ and
$\pi \tilde{I}_2(\alpha=1/4 + i\delta) \equiv R_2 - iP_2  \delta + O(\delta^2)$ where $\delta \ra 0$ regulates an artificial singularity in $\pi \tilde{I}_3(\alpha=1/4)=R_2 + P_2/4$.
With 10000 iterations for each series, we obtain the following numerical estimates:
\be
R_1=163.7428, \quad R_2=209.175, \quad P_2=1260.720 \, .
\label{Is-numerics}
\ee
From these results, we may then obtain the numerical value of $S=R_1-R_2/8-7P_2 /128$ which, combined with the one of $\hat{\Pi}$, yields 
$L_c=32.45$ and therefore $N_c=3.29$. This result shows that the inclusion of the $1/N$ corrections increases the
critical value of $N_c$ by only $1.5\%$ with respect to its LO value.

\section{Conclusion}  

We have included $O(1/N^2)$ contributions to the SD equation exactly and found that the critical value $N_c$ increased
by $1.5\%$ with respect to the LO result.
Our analysis is in nice agreement with ~[\onlinecite{Nash89}] and
therefore gives further evidence in favour of the solution found by Appelquist et al.~[\onlinecite{AppelquistNW88}].
Our results are in support of the fact
that the $1/N$ expansion of the kernel of the SD equation describes reliably the
critical behaviour of the theory.

In closing, let us briefly compare our study with the one of Nash~\cite{Nash89} which, to the best of our knowledge,
is the only popular reference which included NLO contributions in the gap equation of QED$_3$. 
Our good agreement with Ref.~[\onlinecite{Nash89}] is nice but rather strange because
the two analyses are done in quite different ways. While we have used the Landau gauge (in accordance with recent results~\cite{BashirRSR09} 
showing the gauge invariance of $N_c$ in this gauge when using the Ball-Chiu vertex), Nash worked with an arbitrary
gauge fixing parameter, $\xi$. He has resummed the most important NLO terms ($\propto \beta^2$
in our definition) which, together with the LO ones, lead to a gauge invariant result for $N_c$.
This result is larger by a factor $4/3$ than the pure LO one.~\cite{AppelquistNW88} The
rest of the NLO terms ($\propto \beta $) were evaluated (mostly numerically) in the Feynman gauge, which
modifies $N_c$ another time and gives the final result of Nash: $N_c=3.28$.
Finally, we also note that Nash obtained two possible solutions (one was considered as unphysical) while 
we obtained a unique one.
For these reasons, and despite the surprising closeness of the final results, 
our analysis substantially differs from that of Nash and intermediate expressions are difficult to compare.
\footnote{We have been informed by V.~Gusynin that the situation looks even more tricky; Eq.~(15) in Ref.~[\onlinecite{Nash89}] contains an error: 
``341'' should be replaced by ``277'' which then leads to $N_c=3.52$.}

We also note that, very recently, NLO corrections were computed by Gusynin and Pyatkovskiy~\cite{Gusynin:2016som}
using a slightly different approach
than ours; they obtained a gauge-independent value $N_c =2.85$. Their value is remarkably close to the one 
recently obtained by Herbut,~\cite{Herbut:2016ide} $N_c = 2.89$ using a completely different method. 
In order to clear up the beautiful agreement we have with 
Nash's results~\cite{Nash89} as well as the difference with the results of Gusynin and Pyatkovskiy,~\cite{Gusynin:2016som} 
we plan to take into account of all $\xi$-dependent terms in our forthcoming publication.

\acknowledgments
We are grateful to Valery Gusynin for discussions.
One of us (A.V.K.) was supported by RFBR grant 16-02-00790-a.
Financial support from Universit\'e Pierre et Marie Curie and CNRS is acknowledged.

\end{fmffile}

\end{document}